\documentclass[trackchanges,twocolumn]{aastex7}

\usepackage{amsmath}

\begin{document}

\title{Pulsation Mode Identification and Classification of 46 High-Amplitude $\delta$ Scuti Stars from 50 Candidates with TESS observations}

\author[orcid=0000-0002-1859-4949]{Taozhi Yang}
\affiliation{Ministry of Education Key Laboratory for Nonequilibrium Synthesis and Modulation of Condensed Matter, School of Physics, Xi'an Jiaotong University, 710049, People's Republic of China}
\email[show]{yangtaozhi2018@163.com}  

\author[orcid=0000-0001-6693-586X]{Zhaoyu Zuo} 
\affiliation{Ministry of Education Key Laboratory for Nonequilibrium Synthesis and Modulation of Condensed Matter, School of Physics, Xi'an Jiaotong University, 710049, People's Republic of China}
\email[show]{zuozyu@xjtu.edu.cn}

\begin{abstract}

Asteroseismic modelling of high-amplitude $\delta$ Scuti (HADS) variables critically depends on their accurate classification, which provides robust constraints on stellar physical parameters. As a foundational step in this direction, we present a detailed analysis of the pulsational behavior of 50 HADS star candidates using high-precision photometric data from the Transiting Exoplanet Survey Satellite (TESS). We confirm 46 as genuine HADS variables, with 40 stars having their dominant frequencies identified for the first time. Among them, 7 pulsate solely in the fundamental mode; 21 exhibit the fundamental mode alongside at least one low-amplitude nonradial mode; 5 are pure double-mode pulsators (the fundamental and first-overtone modes), 13 show double-mode pulsations accompanied by additional low-amplitude nonradial modes. The remaining four stars are classified as other types of variables: two (TIC 69546708 and TIC 110937533) are confirmed hot subdwarfs, one (TIC 8765832) is a cataclysmic variable, and one (TIC 32302937) is a likely hot subluminous star, but it requires further spectroscopic confirmation. We investigate the period-luminosity (PL) relation, also known as the Leavitt law, for these 46 confirmed HADS stars, deriving a revised relation: $M_{V}= (-3.31 \pm0.39)~\mathrm{log}~P+(-1.68 \pm 0.39)$. This result is consistent with previous studies. Their distribution in the Hertzsprung-Russell diagram indicates that HADS stars are not strictly confined to be within a narrow instability strip previously found, but can extend beyond it, with a distribution toward lower temperatures. The refined classifications presented here establish a high-quality sample for precise asteroseismic modelling and enhance the potential for future machine-learning-assisted searches and classifications of HADS stars in large-scale photometric surveys.

\end{abstract}


\section{Introduction} 

Stars are fundamental building blocks of the universe, responsible for creating elements and shaping galaxies. Gaining a precise understanding of stellar structure and evolution is crucial for uncovering the history of the universe, as well as the formation of galaxies and exoplanets \citep{2018ASSP...49....3S}. Astroseismology, by utilizing the stellar oscillations, can reveal the internal structure of stars and obtain their evolutionary characteristics \citep{2010aste.book.....A,2013ARA&A..51..353C,2015pust.book.....C,2021RvMP...93a5001A}. Stellar oscillations can be observed photometrically with ground-based or space-based optical telescopes \citep{2013pss4.book..207H,2021RvMP...93a5001A} and pulsation frequencies can be obtained by calculating the amplitude spectrum (e.g. \citealt{2005CoAst.146...53L}). Then, by combining stellar pulsation theory and evolutionary models, key stellar parameters such as mass, age, and evolutionary state can be determined. This is particularly valuable because directly observing these properties is often challenging.

Among pulsating variables, asteroseismological studies of $\delta$ Scuti ($\delta$ Sct) variables have great potential. $\delta$ Sct star is a type of short period pulsating variable with periods between 15 min and 8 h, located at the intersection of the classical instability strip and the main sequence \citep{2011A&A...534A.125U,2014MNRAS.439.2078H}. Most $\delta$ Sct stars are multiperiodic pulsators showing radial ($l = 0$) and non-radial ($l > 0$) pulsations, driven by the $\kappa$ mechanism operating in the He-II partial ionisation zone \citep{2000ASPC..210....3B,2010aste.book.....A,2014ApJ...796..118A,2020MNRAS.498.4272M}. They are intermediate-mass Population I stars with masses ranging from 1.5 to 2.5 M$_{\odot}$, occupying the transition region between lower-mass stars with thick outer convective envelopes and higher-mass stars with thin convective shells \citep{2021MNRAS.504.4039B}. Consequently, the pulsations of $\delta$ Sct stars provide a valuable means to probe the internal structure and evolutionary state of stars within this transitional regime with high precision.

Within the broader class of $\delta$ Sct star, the subgroup of high-amplitude $\delta$ Scuti (HADS) stars is typically characterized by large photometric amplitude ($\Delta$V $\ge$ 0.1 mag) and slow rotation ($v$ sin i $\leq$ 30 km/s). Their pulsation periods range from approximately 1 to 6 hrs \citep{2000ASPC..210..373M}. Most HADS variables are either single-mode pulsators (the fundamental mode only) or double-mode variables exhibiting both fundamental and first overtone modes \citep{1997ApJ...477..346B,2008A&A...478..865W,2011AJ....142..110M,2021AJ....161...27Y,2022MNRAS.515.4574N,2023ApJ...959...33L,2023RAA....23g5002X}. The advent of large-scale surveys has led to the discovery of an increasing number of multi-mode HADS variables, with some stars showing three or even four simultaneously excited radial modes \citep{2016AJ....152...17M,2022ApJ...932...42L}. These provide valuable opportunities for a more precise determination of the internal structure of stars. In addition to radial pulsation modes, non-radial pulsation modes have been detected in some HADS variables, as well \citep{2011A&A...534A.125U}.

In recent years, studies of HADS stars have advanced dramatically, owing to the successful operation of a series of space observation missions, such as MOST \citep{2003PASP..115.1023W}, CoRoT \citep{2009A&A...506..411A}, Kepler \citep{2010Sci...327..977B},  and TESS \citep{2014SPIE.9143E..20R}. The advent of high-precision space-based photometry has revolutionized the study of variable stars, allowing the detection of low-amplitude oscillations and yielding a wealth of new discoveries. Using 4-yrs of observations from the $Kepler$ telescope, \cite{2018ApJ...863..195Y} found a pair of low-amplitude triplet structures in the frequency spectra of HADS star KIC 5950759 and the reason for this triplet structure might be the amplitude modulation due to stellar rotation. In another HADS star KIC 10284901, a weak amplitude modulation with two frequencies was also discovered, which may provide new observational evidence for the Blazhko effect \citep{2019ApJ...879...59Y}. Meanwhile, the modulation of the amplitude of $\delta$ Sct star can also be systematically explored thanks to the long time-span data from $Kepler$ mission. \cite{2016MNRAS.460.1970B} conducted a systematic search for pulsation amplitude modulation in 983 $\delta$ Sct stars continuously observed by the Kepler telescope for four years. They found that more than half of the sample exhibit at least one pulsation mode that varies significantly in amplitude over the observation period. In addition, a slow decline in amplitude of the first overtone mode in HADS star KIC 2857323 was first detected by \cite{2022ApJ...936...48Y} and they suggested that the amplitude decline in this star may be due to pulsation energy loss$-$a mechanism requiring further investigation to confirm its physical nature.

Despite progress in characterizing HADS variables, a key unresolved question remains: why do most exhibit single-period pulsations, while only a minority show double-mode or multi-mode behavior? Furthermore, what differences exist in physical parameters (e.g. mass, luminosity, and metallicity) and evolutionary states between these groups? Addressing these questions would significantly advance our understanding of the pulsation mechanism and stellar internal structure. Expanding the sample of well-classified HADS stars through detailed studies will be crucial for making progress on these aforementioned issues. Fortunately, the nearly all-sky coverage of TESS provides an excellent opportunity to address this goal. Based on observations from the first 67 sectors, \citet{2025ApJS..276...57G} identified 72,505 periodic variable stars, including 63,106 new discoveries. Using a random forest classification method, they categorized these newly identified variables into 12 subtypes, among which 50 were classified as HADS candidates -- substantially expanding the known population of such objects. However, their work did not include detailed verification of the nature of these candidates or further pulsational mode classifications.

To address the need for refined classifications and build a high-quality sample for future studies, we perform a detailed analysis of these HADS candidates. Our study begins with an introduction to the TESS observations and data reduction procedures in Section 2. We then present the frequency analysis and mode identification for the dominant frequencies in Section 3, which forms the basis for the classification results detailed in Section 4. Building upon these results, we investigate the PL relation (Leavitt Law) of the sample in Section 5 and place the stars in the Hertzsprung-Russell diagram to discuss their evolutionary stages in Section 6. The broader implications of our findings are discussed in Section 7, and the study concludes with a summary in Section 8.

\section{TESS observations and Data Reduction} \label{sec:sobservations}

While the primary goal of the TESS mission is exoplanet discovery and characterization \citep{2014SPIE.9143E..20R}, its high-precision photometric data also offer significant opportunities for stellar astrophysics and asteroseismology. It mainly provides two types of photometric measurements for stars: 2-min cadence light curves and full frame images every 30 minutes. There are also some targets that are observed by 20 seconds and 200 seconds cadence. All these light curves can be accessed through the Mikulski Archive for Space Telescopes (MAST)\footnote{\url{https://archive.stsci.edu/}}. The specific data used in this study were obtained from the "TESS Data For Asteroseismology Lightcurves (TASOC)" \citet{https://doi.org/10.17909/t9-4smn-dx89} repository at MAST. To conduct a more detailed study of pulsating behavior and accurate classification of these 50 HADS candidates, we primarily utilize 2-min cadence photometric data processed by the TESS Science Processing Operations Center (SPOC) \citep{2016SPIE.9913E..3EJ}, using the Lightkurve package \citep{2018ascl.soft12013L}. The 2-min cadence photometric data ensure that we detect all the high-frequency oscillations, especially beneficial for pulsators like $\delta$ Sct-type variables. Light curves obtained from multiple TESS sectors also provide a longer time baseline which ensures a good accuracy on the detected frequencies. 

\begin{figure}
  \centering
\includegraphics[width=0.45\textwidth,trim=30 245 25 245,clip]{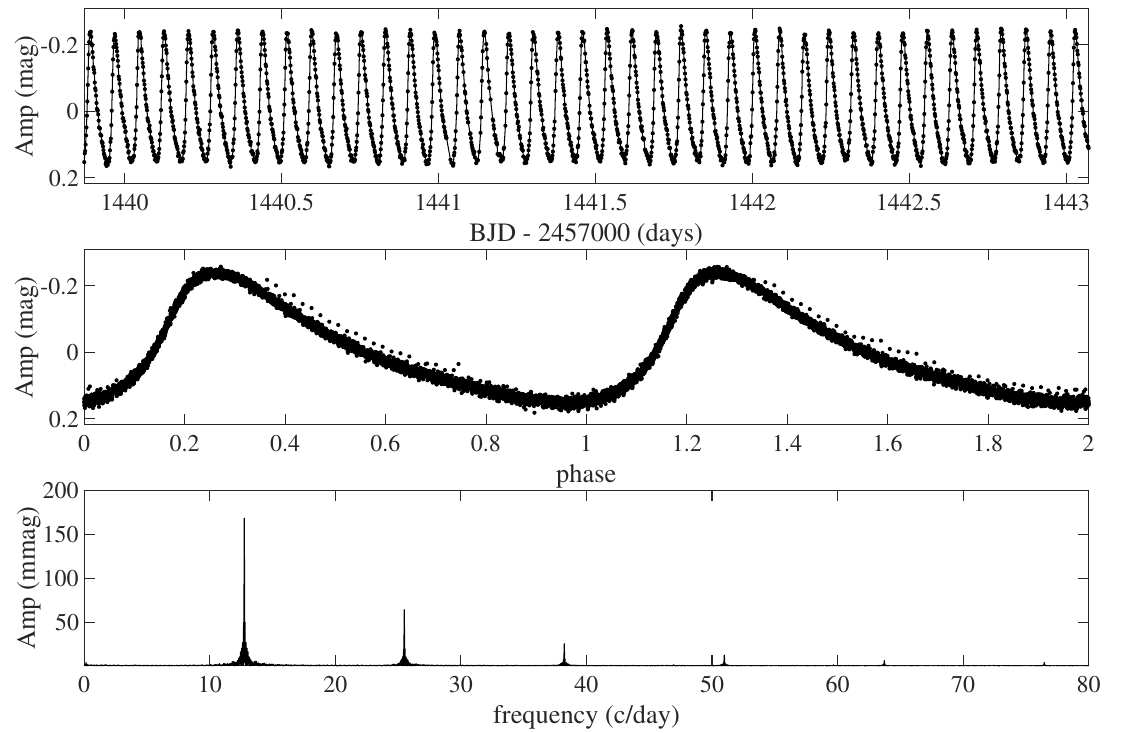}
\caption{Top pannel: A portion of light curve of TIC 710783 for about 3 days; Middle pannel: Phased diagram (folded by the dominant frequency $f = 12.74450 $ c/day); Lower pannel: Amplitude spectra with a frequency range of $0 < f < 80$ c/day.}
\label{fig:Fig1_light_curve}
\end{figure}

In this work, we use the Pre-search Data Conditioning (PDC)-SAP fluxes of these stars. For each quarter, we first remove the obvious outliers and de-trend the light curves with linear or second-order polynomial. Then, the flux data are converted to the magnitude scale, and each quarter is adjusted to zero by subtracting its mean value. Finally, all quarters are stitched to a total light curve. As an illustrative example, Figure \ref{fig:Fig1_light_curve} presents a 3-day segment of the light curve for TIC 710783, along with its phased diagram folded at the dominant frequency ($f = 12.74450$ c/day). The phased light curve reveals characteristic HADS variability with a peak-to-peak amplitude of $\sim$0.45 mag, while the corresponding amplitude spectra displays the relatively simple harmonic structure typical of HADS pulsators.

\section{Frequency Analysis and Mode Identification} \label{sec:fre_analysis}

The pulsation behavior of the candidate stars was investigated through frequency extraction using the software PERIOD04\citep{2005CoAst.146...53L}, in which the light curve is fitted with the following formula:

\begin{equation}
m = m_{0} + \Sigma\mathnormal{A}_{i}sin(2\pi(\mathnormal{f}_{i}\mathnormal{t} + \phi_{i}))
\end{equation}
 
where $m_{0}$ is the zero-point, $A_{i}$, $f_{i}$, and $\phi_{i}$ represent the amplitude, frequency, and the corresponding phase, respectively.

To detect more potential pulsation modes, we chose a wider frequency range of $0 < f < 80$ c/day, which covers the typical pulsation frequency range of $\delta$ Sct stars. The peaks were extracted one-by-one via the standard method of prewhitening, as adopted by \cite{2019ApJ...879...59Y}. For determining the significance of a detected peak, we employed a signal-to-noise ratio (S/N) criterion with a threshold of 5.2, following the recommendation of \cite{2015MNRAS.448L..16B}. The uncertainty in frequency was determined using the method proposed by \cite{1999DSSN...13...28M}, allowing us to estimate the accuracy of frequency measurements.

In HADS stars, the highest-amplitude pulsation frequency is typically identified as the fundamental radial mode. We adopt this convention, noting that the first overtone can dominate in some cases \citep[e.g. HD 224852;][]{2005A&A...440.1097P}. Once the fundamental mode is established, subsequent strong frequencies are assessed as potential radial overtones using their period ratios against the fundamental mode, following the criteria from \citet{1979ApJ...227..935S}:
\begin{equation}
\begin{split}
  0.756 & \leq P_{1}/P_{0}  \leq 0.787  \\
  0.611 & \leq P_{2}/P_{0}  \leq 0.632  \\
  0.500 & \leq P_{3}/P_{0}  \leq 0.525 
\end{split}
\end{equation}

where $P_{0}$, $P_{1}$, $P_{2}$, and $P_{3}$ represent the periods of the fundamental, first overtone, second overtone, and third overtone pulsation modes, respectively. In this work, we followed the relations in Equation (1) to perform the mode identification for the radial mode. If a frequency meets the relation: $f_{i}=m*f_{j} \pm n*f_{k}$ ($f_{j}$ and $f_{k}$ are parent frequencies, usually with higher amplitude, $m$ and $n$ are small integer) within the frequency resolution of 1.5/$\Delta$T \citep{1978Ap&SS..56..285L}, then it would be judged as a combination of the dominant frequencies. Besides, the lower frequencies that neither meet the period ratios in Equation (1) nor are combinations would be considered as non-radial pulsation modes. The identification of non-radial pulsation modes is usually complex and therefore not the focus of this work. For convenience, the fundamental mode, first overtone and second overtone are abbreviated as F, 1O, and 2O respectively in this work.

\section{Results}

Based on the detected frequencies and their period ratios, these 50 HADS candidates were grouped into five distinct subtypes. These include: mono-mode HADS stars (Table \ref{tab:list of single-mode HADS}), HADS stars with F mode and non-radial modes (Table \ref{tab:list of HADS with fundamental mode and non-radial mode}), double-mode HADS stars (Table \ref{tab:list of pure double-mode HADS}), double-mode HADS stars with non-radial modes (Table \ref{tab:list of double-mode HADS with non-radial mode}), and other types of variable stars (Table \ref{tab:list Other types of variable stars}). Each table includes IDs, dominant frequencies, periods, amplitudes, and TESS sectors. Full frequency sets appear in Table \ref{tab:TIC710783}, while the light curves and the amplitude spectra are provided in the supplementary material (Lightcurve).

\subsection{Mono-mode HADS stars}

Mono-mode HADS stars are defined as high-amplitude $\delta$ Sct variables exhibiting pulsations dominated by a single intrinsic frequency, typically corresponding to the fundamental radial mode. It is relatively easy to identify the mono-mode HADS star. In their amplitude spectra, the strongest peak was considered as the frequency of the fundamental mode, and all others are its harmonics. After prewhitening all these frequencies, the residuals of remaining spectra show no significant frequencies. In total, 7 stars are found to be mono-mode HADS stars in this sample, as listed in Table \ref{tab:list of single-mode HADS}. From the table, it seems that the majority of their pulsation frequencies are less than 10 c/day, which may represent a pulsating feature of mono-mode HADS stars. 

\begin{deluxetable}{rrrccc}
\tablewidth{0pt}
\tablecaption{List of 7 HADS stars with only F mode
\label{tab:list of single-mode HADS}}
\tablehead{
\colhead{ID} & \colhead{TIC number} & \colhead{Frequency } & \colhead{Period } & \colhead{Amplitude} & \colhead{Sector(s)} \\ 
\colhead{} & \colhead{} & \colhead{(c/day)} & \colhead{ (day) } & \colhead{ (mmag) }
}
\startdata
1 &   9632550   &   5.05505  &  0.19782  &  121.78   & 29  \\
2 &  17153995   &   8.54480  &  0.11703  &  168.51   & 12  \\
3 &  18426988   &   8.56578  &  0.11674  &   50.04   & 35  \\     
4 &  60334458   &   7.60801  &  0.13144  &   78.93   & 43, 44  \\
5 &  62628148   &   8.29033  &  0.12062  &   62.72   & 34  \\
6 & 120472041   &  12.61482  &  0.07927  &   69.60   & 38  \\
7 & 138735041   &   8.65753  &  0.11551  &   99.52   & 30  \\
\enddata
\end{deluxetable}

\begin{figure}
  \centering
\includegraphics[width=0.45\textwidth,trim=30 245 25 245,clip]{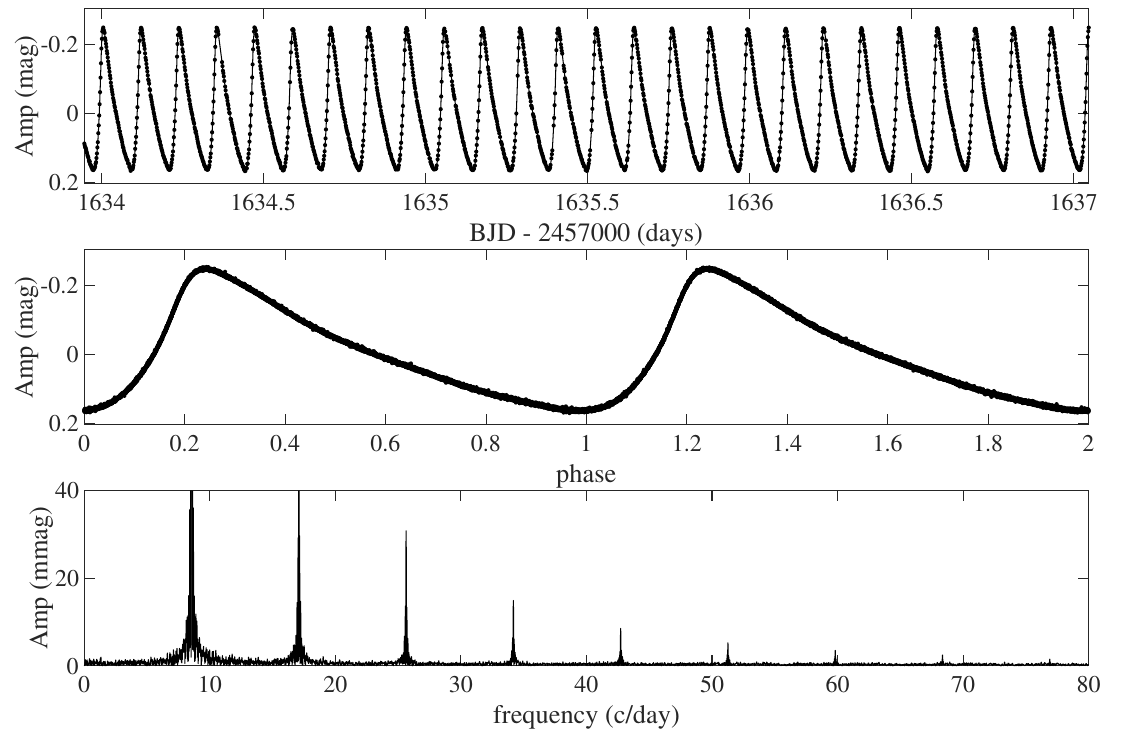}
\caption{The upper panel shows a portion of the light curve of TIC 17153995 for about 3 days; The middle panel shows the phase diagram folded by the fundamental frequency $f = 8.54480$ c/day; The lower panel shows the Fourier amplitude spectra.}
\label{fig:TIC_17153995_light_curve}
\end{figure}

Figure \ref{fig:TIC_17153995_light_curve} shows the light curves and the amplitude spectra of TIC 17153995 as a typical example of mono-mode HADS star. It is clear that the light curve and phase diagram clearly exhibit the characteristic rapid-rise and slow-fall pattern unique to HADS variables, while the amplitude spectra confirms its mono-mode nature through a simple, single frequency structure.

\subsection{HADS stars with F mode and non-radial modes}

Among all detected frequencies, we identified several that cannot be explained as fundamental-mode harmonics or radial-mode combinations (based on period ratios). They are labbled as non-radial pulsation modes. We classified 21 such  HADS stars into this subgroup (Table \ref{tab:list of HADS with fundamental mode and non-radial mode}), characterized by not only a dominant fundamental radial frequency, but also one or more additional non-radial frequencies. Their frequency distribution shows significantly broader ranges ($5 < f < 15$ c/day) than mono-mode HADS variables, suggesting more complex pulsation behavior.

\begin{deluxetable}{rrrrrc}
\tablewidth{0pt}
\tablecaption{List of 21 HADS stars with F mode and non-radial modes
\label{tab:list of HADS with fundamental mode and non-radial mode}}
\tablehead{
\colhead{ID} & \colhead{TIC number} & \colhead{Frequency} & \colhead{Period} & \colhead{Amplitude} & \colhead{Sectors} \\ 
\colhead{} & \colhead{} & \colhead{(c/day)} & \colhead{(day)} & \colhead{(mmag)}
}
\startdata
 1  &     710783  &  12.74450  &  0.07847  & 168.97  &  5  \\
 2  &    7082633  &  10.32293  &  0.09687  &  50.66  & 13  \\
 3  &   16283570  &  11.02399  &  0.09071  & 117.26  & 50, 51  \\
 4  &   16381609  &   5.56039  &  0.17984  &  75.06  & 71  \\
 5  &   27315670  &   9.15190  &  0.10927  & 124.80  & 14, 15, 16 \\
 6  &   28292974  &  14.27855  &  0.07004  &  62.97  & 17  \\
 7  &   34137913  &  13.13146  &  0.07615  & 112.95  &  6  \\
 8  &   35166984  &  10.48515  &  0.09537  & 171.72  &  6  \\
 9  &   37752140  &  12.73171  &  0.07854  &  67.14  & 42  \\
10  &   51991595  &  10.91561  &  0.09161  & 131.63  &  1  \\
11  &   60405689  &  14.98543  &  0.06673  &  62.00  & 33  \\
12  &   67265166  &   7.34619  &  0.13612  &  63.11  & 87  \\
13  &   78850814  &   9.11832  &  0.10967  &  74.74  & 35, 36  \\
14  &   85914188  &   8.28486  &  0.12070  & 168.63  & 45, 46  \\
15  &   90185615  &   7.20049  &  0.13888  &  66.26  & 13  \\
16  &  112682462  &   9.13442  &  0.10948  & 187.54  & 39  \\
17  &  129128212  &  10.49790  &  0.09526  &  97.68  & 16  \\
18  &  131351117  &   8.75820  &  0.11418  & 134.22  & 37  \\
19  &  137803552  &  10.94784  &  0.09134  & 125.25  & 21  \\
20  &  139845816  &   6.80213  &  0.14701  & 156.01  & 28  \\
21  &  141457913  &   9.45734  &  0.10574  &  94.59  &  7  \\
\enddata
\end{deluxetable}

\begin{figure}
  \centering
\includegraphics[width=0.45\textwidth,trim=30 245 25 245,clip]{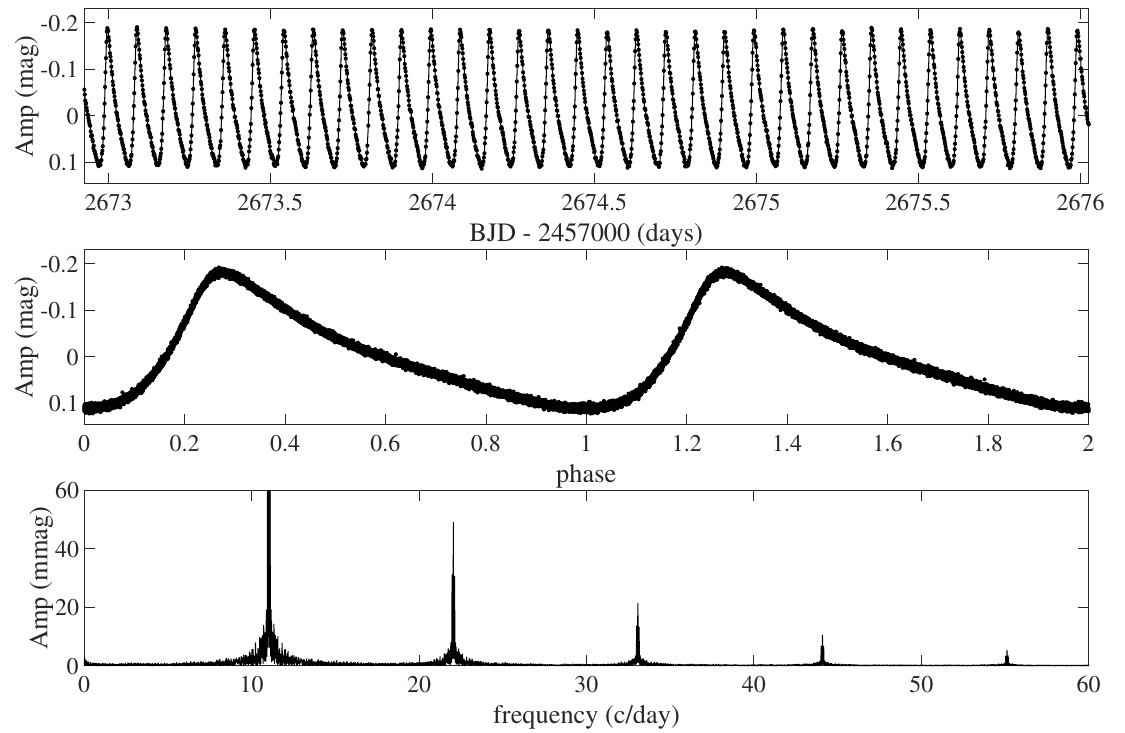}
\caption{The upper panel shows a portion of the light curve of TIC 16283570 for about 3 days; The middle panel shows the phase diagram folded by the fundamental frequency $f = 11.02399$ c/day; The lower panel shows the Fourier amplitude spectra.}
\label{fig:TIC_16283570_light_curve}
\end{figure}

\begin{figure}
  \centering
\includegraphics[width=0.45\textwidth,trim=30 245 25 245,clip]{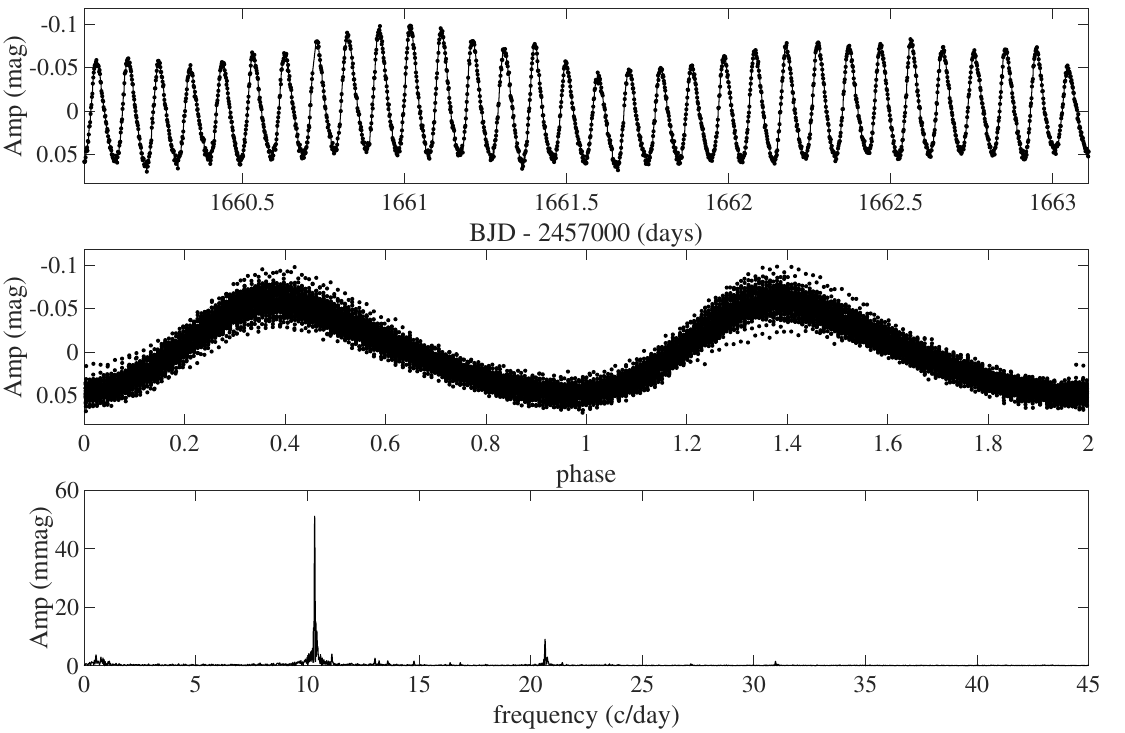}
\caption{The upper panel shows a portion of the light curve of TIC 7082633 for about 3 days; The middle panel shows the phase diagram folded by the fundamental frequency $f = 10.32293$ c/day; The lower panel shows the Fourier amplitude spectra.}
\label{fig:TIC_7082633_light_curve}
\end{figure}

Figure \ref{fig:TIC_16283570_light_curve} shows the light curves and the amplitude spectra of TIC 16283570 as a typical example of this group of variable star. From this figure, it is clear that its light curve is very similar to that of the mono-period HADS variables, because its light variation is also dominated by the fundamental mode, and the amplitude of the non-radial mode is very weak, which is less than 0.6 \% of the amplitude of the fundamental mode. However, the light curves and the amplitude spectra of TIC 7082633, as shown in Figure \ref{fig:TIC_7082633_light_curve}, show differences from the case of TIC 16283570. This is because its variability arises  not only from the fundamental mode but also from multiple non-radial modes-most notably $f_{3}$ = 10.43013 c/day, whose amplitude (13.6 \% of the fundamental one) includes clear amplitude modulation in the light curve and phase discretization in the folded diagram (see middle panel in Figure \ref{fig:TIC_7082633_light_curve}). These two different types of light curves would be helpful for accurate classification of HADS stars when using machine-learning-assisted methods in the large-field photometric surveys.

Among these stars, TIC 710783 was first discovered to be a variable star by \cite{2018AJ....156..241H}, and subsequently classified as a $\delta$ Sct star based on its physical parameters by \cite{2020A&A...638A..59B}. \cite{2024ApJ...969...19V} analyzed the pulsation of this star with TESS data and proposed that it may be a new mono-mode HADS, which is consistent with the result in this work. TIC 28292974 (another identifiers: BN Tri) were first acquired from the ROTSE-I survey between 1999 and 2000 \citep{2000AJ....119.1901A,2004AJ....127.2436W} and  later from the Catalina Sky \citep{2014ApJS..213....9D} and SuperWASP surveys \citep{2006PASP..118.1407P}. \cite{2007PZP.....7...25K} initially identified this pulsating variable as a $\delta$ Scuti-type star based on data from the ROTSE-I survey. \cite{2022JAVSO..50..218A} analyzed the pulsations of this star using TESS data and detected two other low amplitude independent frequencies besides the fundamental frequency. 

TIC 112682462 was a known $\delta$ Sct star by \cite{1987IzKry..76...10K}, and identified as a HADS star by \cite{2023ApJ...959...33L} using TESS data. They reported that TIC 112682462 might be a double-mode HADS with the fundamental mode (F=9.13442 c/day) and a third overtone mode ($f_{3}$=16.96056 c/day). However, considering that the ratio of F/$f_{3}$ (=0.5386) is larger than the theoretical ratio (0.500-0.525), we propose that the frequency $f_{3}$ should belong to a non-radial mode, instead. 

\subsection{Double-mode HADS stars}

Table \ref{tab:list of pure double-mode HADS} lists 5 HADS stars, which can be identified as pure double-mode HADS stars, that is to say, they are pulsating in only the fundamental and first overtone modes, without any other modes. From the table, it is clear that the period ratio of F/1O is in a very narrow range, namely 0.77-0.78. 

Among these stars, the first 4 stars (TIC 17539022, TIC 56914404, TIC 79662102, and TIC 81709032) are newly identified. The last one TIC 130474019 was studied by \cite{2020A&A...638A..59B} using TESS data and its physical parameters (temperature, gravity, and frequency scaling relation) was reported. Subsequently, \cite{2022MNRAS.516.2080B} classified this star as a $\delta$ Sct star. \cite{2024ApJ...969...19V} identified this star as a double-mode HADS star through analysis of its radial-mode frequency ratios using two-sector data from TESS.

\begin{deluxetable*}{rrrrrc}
\tablewidth{0pt}
\tablecaption{List of 5 HADS stars with only F and 1O modes
\label{tab:list of pure double-mode HADS}}
\tablehead{
\colhead{ID} & \colhead{TIC number}  & \colhead{$P_0$}  & \colhead{$P_1$} & \colhead{$P_1/P_0$} & \colhead{Sectors} \\ 
\colhead{} & \colhead{} & \colhead{(day)} & \colhead{(day)} & \colhead{} & \colhead{}
}
\startdata
1  &  17539022  &  0.05172  &  0.03999  & 0.77320 &  51, 52  \\
2  &  56914404  &  0.09587  &  0.07450  & 0.77709 &  21  \\
3  &  79662102  &  0.05982  &  0.04628  & 0.77365 &  27  \\
4  &  81709032  &  0.11157  &  0.08621  & 0.77270 &  34, 35  \\
5  & 130474019  &  0.08142  &  0.06293  & 0.77291 &  07  \\
\enddata
\end{deluxetable*}

\subsection{Double-mode HADS stars with non-radial modes}

\begin{figure}
  \centering
\includegraphics[width=0.45\textwidth,trim=30 245 25 245,clip]{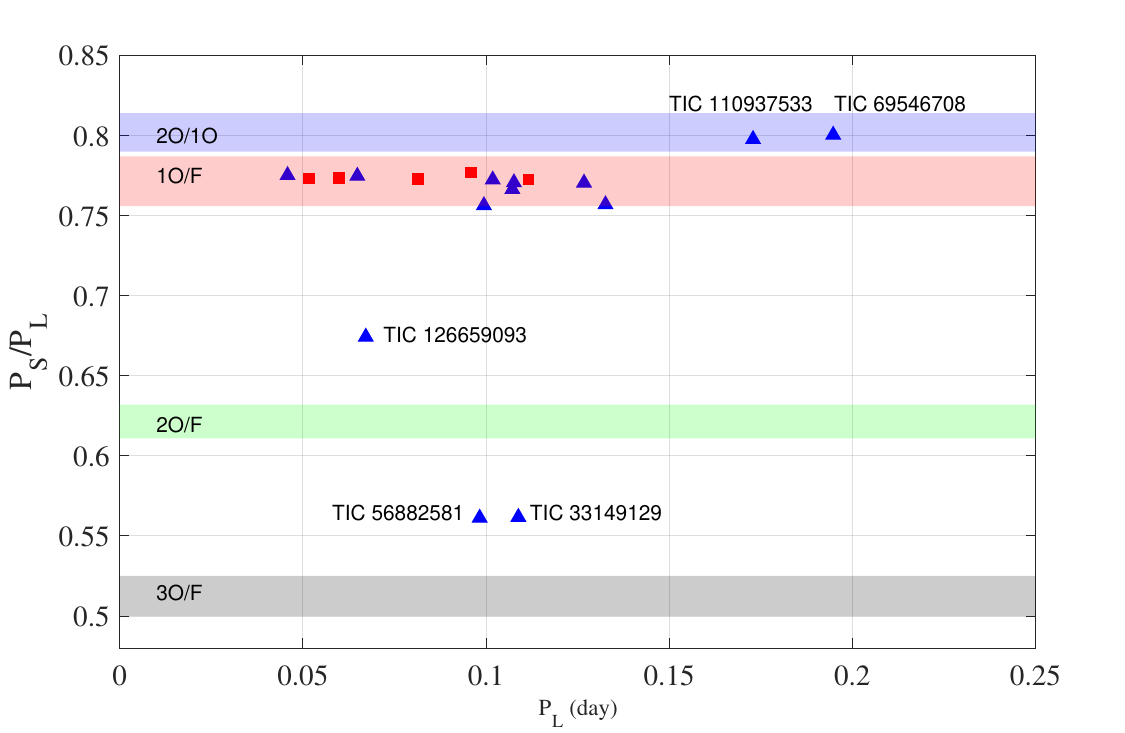}
\caption{The Peterson diagram of the double-mode HADS stars. Red squares represent 5 pure double-mode HADS stars, blue triangles are 13 HADS in this group. The shaded bands represent the period ratios of the first three overtone modes to fundamental mode, as well as the period ratio of the second overtone to first overtone mode \citep{1979ApJ...227..935S}.}
\label{fig:PD_Multi_mode}
\end{figure}

In addition to the fundamental and first overtone mode, some non-radial modes are also detected in HADS stars. Table \ref{tab:list of double-mode HADS with non-radial mode} lists 13 HADS stars, in which they pulsate with both two strong radial modes and several low amplitude non-radial modes. Figure \ref{fig:PD_Multi_mode} presents the Peterson diagram of 18 double-mode HADS stars, including 5 pure double-mode HADS stars (red squares) and 13 double-mode pulsators with additional non-radial modes in this group (blue triangles). It is clear that the period ratio of 8 stars (TIC 25671619, TIC 30542468, TIC 30977864, TIC 46937596, TIC 90322352, TIC 91592810, TIC 97937351, and TIC 118080601) lie in the range of 0.756 $\sim$ 0.787, which indicate the two strong modes belong to the fundamental and first overtone modes. Two stars (TIC 33149129 and TIC 56882581) have a lower period ratio of about 0.56, that do not match any value of the period ratios of the first four radial modes for $\delta$ Sct stars, indicating that the second strong frequency in both of the two stars must belong to the same non-radial modes. The period ratios of two other stars, TIC 69546708 and TIC 110937533, are approximately 0.80—significantly higher than the typical $P_{1}/P_{0}$ range of 0.756 $\sim$ 0.787, but consistent with the theoretical ratio of the second to first overtone  (2O/1O). This suggests that the strongest frequency in both systems likely corresponds to the first overtone mode. The last star, TIC 126659093, exhibits a period ratio of 0.67431, which lies between the typical values for 2O/F and 1O/F, strongly indicating that the secondary frequency represents a non-radial mode. To unambiguously determine the pulsation modes in these atypical cases, detailed asteroseismic modelling will be essential in future studies.

TIC 30977864 was a previously known $\delta$ Sct star reported by \cite{1933AN....249..253G}, and identified as a HADS star by \cite{2023ApJ...959...33L} using TESS data. They analyzed its pulsations and also detected two radial modes (the fundamental mode and first overtone mode) and one non-radial mode, which is consistent with the result in this work. 

TIC 90322352 was a known $\delta$ Sct star by \cite{2022MNRAS.516.2080B}, and identified as a HADS star by \cite{2023ApJ...959...33L} using TESS data. According to the detected frequencies, they claimed that it might be a new triple-mode HADS star, namely, $f_{1}$ = 9.32892 c/day is the fundamental frequency, $f_{2}$ = 12.17120 c/day is the first overtone frequency, and $f_{12}$ = 15.01271 c/day is the second overtone mode. However, we propose that $f_{12}$ is not the second overtone frequency. It is actually the combination of $f_{1}$ and $f_{2}$, namely, $f_{12}$ = 2$f_{2}$-$f_{1}$ =15.01348 c/day within the frequency resolution (=0.059 c/day). Therefore, TIC 90322352 is a double-mode HADS star with non-radial mode.

\begin{deluxetable*}{rrrrrc}
\tablewidth{0pt}
\tablecaption{List of 13 HADS stars with two strong modes and non-radial modes
\label{tab:list of double-mode HADS with non-radial mode}}
\tablehead{
\colhead{ID} & \colhead{TIC number}  & \colhead{$P_0$}  & \colhead{$P_1$} & \colhead{$P_1/P_0$} & \colhead{Sectors} \\ 
\colhead{} & \colhead{} & \colhead{(day)} & \colhead{(day)} & \colhead{} & \colhead{}
}
\startdata
 1  &   25671619   &  0.09942  &  0.07520  &  0.75639  &  08  \\
 2  &   30542468   &  0.04582  &  0.03552  &  0.77521  &  37  \\
 3  &   30977864   &  0.10762  &  0.08294  &  0.77070  &  27  \\
 4  &   33149129   &  0.10883  &  0.06114  &  0.56179  &  33  \\
 5  &   46937596   &  0.12675  &  0.09766  &  0.77049  &  12  \\
 6  &   56882581   &  0.09827  &  0.05516  &  0.56131  &  21  \\
 7  &   69546708   &  0.19477  &  0.15590  &  0.80043  &  40, 41  \\
 8  &   90322352   &  0.10719  &  0.08216  &  0.76649  &  13  \\
 9  &   91592810   &  0.06487  &  0.05026  &  0.77478  &  30, 31  \\
10  &   97937351   &  0.10183  &  0.07866  &  0.77246  &  31  \\
11  &  110937533   &  0.17289  &  0.13792  &  0.79773  &  54  \\
12  &  118080601   &  0.13261  &  0.10038  &  0.75696  &  8, 9, 10  \\
13  &  126659093   &  0.06718  &  0.04530  &  0.67431  &  27  \\
\enddata
\end{deluxetable*}

\subsection{Other types of variable stars}

Four stars from the initial candidate list were reclassified as other types of variable stars (listed in Table \ref{tab:list Other types of variable stars}), based on distinct characteristics in their light curves and extracted frequencies that differ markedly from those of HADS stars.

\begin{deluxetable*}{rrrrrc}
\tablewidth{0pt}
\tablecaption{List of other types of variable stars
\label{tab:list Other types of variable stars}}
\tablehead{
\colhead{ID} & \colhead{TIC number}  & \colhead{Period } & \colhead{Amplitude} & \colhead{Sectors} \\ 
\colhead{} & \colhead{} & \colhead{(day)} & \colhead{(mmag)} & \colhead{}
}
\startdata
1 &   1973623   &  0.12675  &  79.58  &  88, 89  \\
2 &   8765832   &  0.07868  &  48.68  &  21  \\
3 &  32302937   &  0.26173  &  86.04  &  88  \\
4 &  36736263   &  0.07299  &  49.88  &  89  \\
\enddata
\end{deluxetable*}

\begin{figure}
  \centering
\includegraphics[width=0.45\textwidth,trim=30 245 25 245,clip]{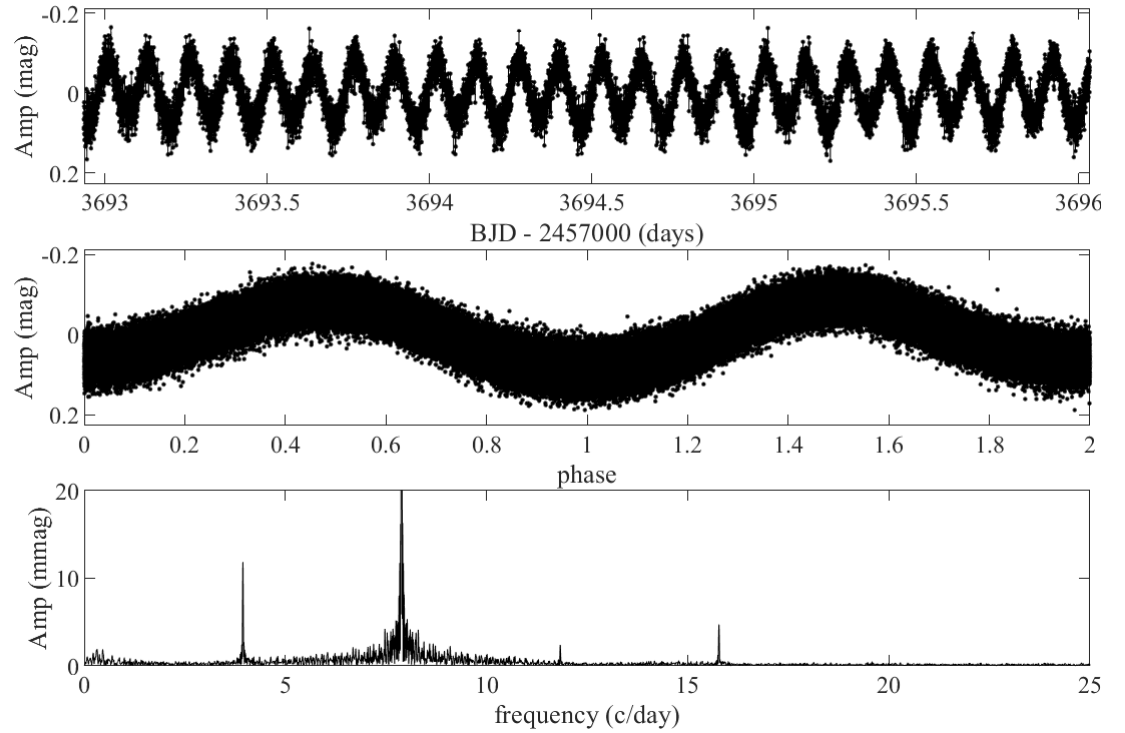}
\caption{The upper panel shows a portion of the light curve of TIC 1973623 for about 3 days; The middle panel shows the phase diagram folded by the strongest frequency $f = 7.88921$ c/day; The lower panel shows the Fourier amplitude spectra.}
\label{fig:TIC_1973623_light_curve}
\end{figure}

Figure \ref{fig:TIC_1973623_light_curve} shows the light curves and the amplitude spectra of TIC 1973623. From this figure, it is clear that the shape of its light curve is very symmetrical within one cycle, contrasting sharply with the characteristic of fast-rise and slow-decline morphology typical of HADS variables. In fact, TIC 1973623 was first reported as a candidate of hot subdwarf star in a southern all-sky hot subdwarf variability survey by \cite{2020ApJ...890..126R}, and subsequently classified as a close binary with a remnant stellar core and an unseen white dwarf (WD) companion by \cite{2020ApJ...902...92R}. Their analysis reveals that the primary of this system is potentially an inflated hot subdwarf (sdO) and more likely is a rarer post-blue horizontal branch (post-BHB) star, thus providing a glimpse into a brief phase of remnant core evolution and secondary variation not seen before in a compact binary. 

\begin{figure}
  \centering
\includegraphics[width=0.45\textwidth,trim=30 245 25 245,clip]{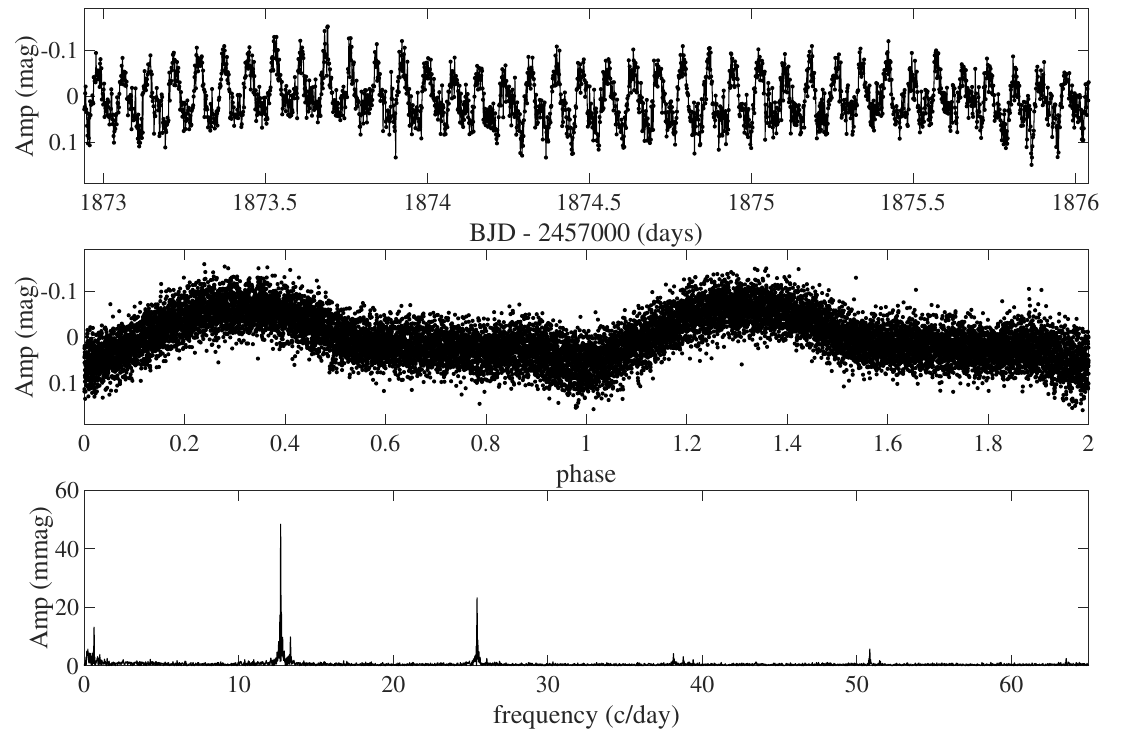}
\caption{The upper panel shows a portion of the light curve of TIC 8765832 for about 3 days; The middle panel shows the phase diagram folded by the strongest frequency $f = 12.70964$ c/day; The lower panel shows the Fourier amplitude spectra.}
\label{fig:TIC_8765832_light_curve}
\end{figure}

For star TIC 8765832, its light curve is significantly different from that of HADS star, mainly manifested in a flat situation in the latter half of a phase cycle, as shown in the middle panel of Figure \ref{fig:TIC_8765832_light_curve}. TIC 8765832 (another identifier: BK Lyn), was discovered as an object with ultraviolet (UV) excess in the Palomar-Green survey \citep{1986ApJS...61..305G}, and classified as a candidate of cataclysmic variable (CV) star by \cite{1982PASP...94..560G}. A subsequent radial-velocity observations confirmed the CV identification and revealed an orbital period of Porb = 107.97 min \citep{1996MNRAS.278..125R}. \cite{2013MNRAS.434.1902P} reported a transition from nova-like to dwarf nova for this star, the possible remnant of the possible classical nova Nova Lyn 101. Recent research on TIC 8765832 mainly focus on the long-term activity and show that it lies close to the upper limit of the luminosity in which dwarf nova (DN) outbursts occur \citep{2024AJ....167..152S}.

\begin{figure}
  \centering
\includegraphics[width=0.45\textwidth,trim=30 245 25 245,clip]{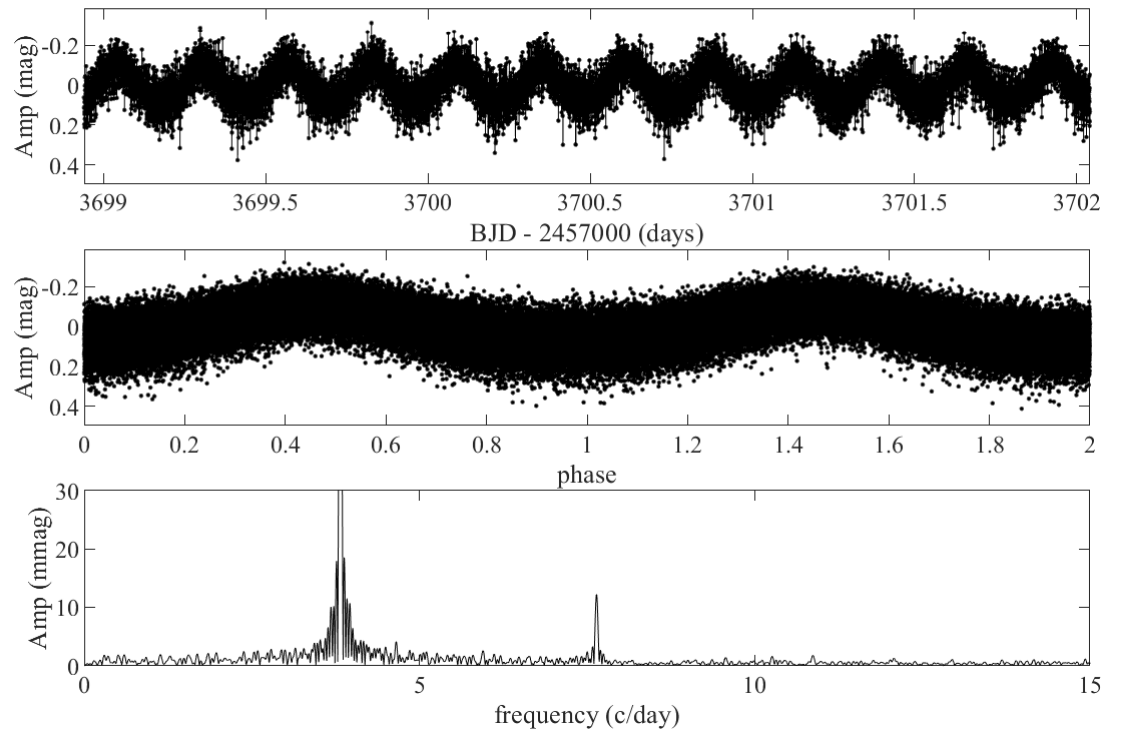}
\caption{The upper panel shows a portion of the light curve of TIC 32302937 for about 3 days; The middle panel shows the phase diagram folded by the strongest frequency $f = 3.82072$ c/day; The lower panel shows the Fourier amplitude spectra.}
\label{fig:TIC_32302937_light_curve}
\end{figure}

Figure \ref{fig:TIC_32302937_light_curve} shows the light curves and amplitude spectra of TIC 32302937, revealing only a dominant frequency and its harmonic. This simple spectrum structure differs markedly from the mono-mode pulsations typical seen in HADS stars. Actually, TIC 32302937 was first discovered as a pulsating star in the Asteroid Terrestrial-impact Last Alert System (ATLAS) by \cite{2018AJ....156..241H}, and subsequently listed as a hot subluminous star candidates based on data from the ESA Gaia Data Release 2 (DR2) and several ground-based, multi-band photometry surveys \citep{2019A&A...621A..38G}. Note that the future spectral observations are necessary for exploring its nature.

\begin{figure}
  \centering
\includegraphics[width=0.45\textwidth,trim=30 245 25 245,clip]{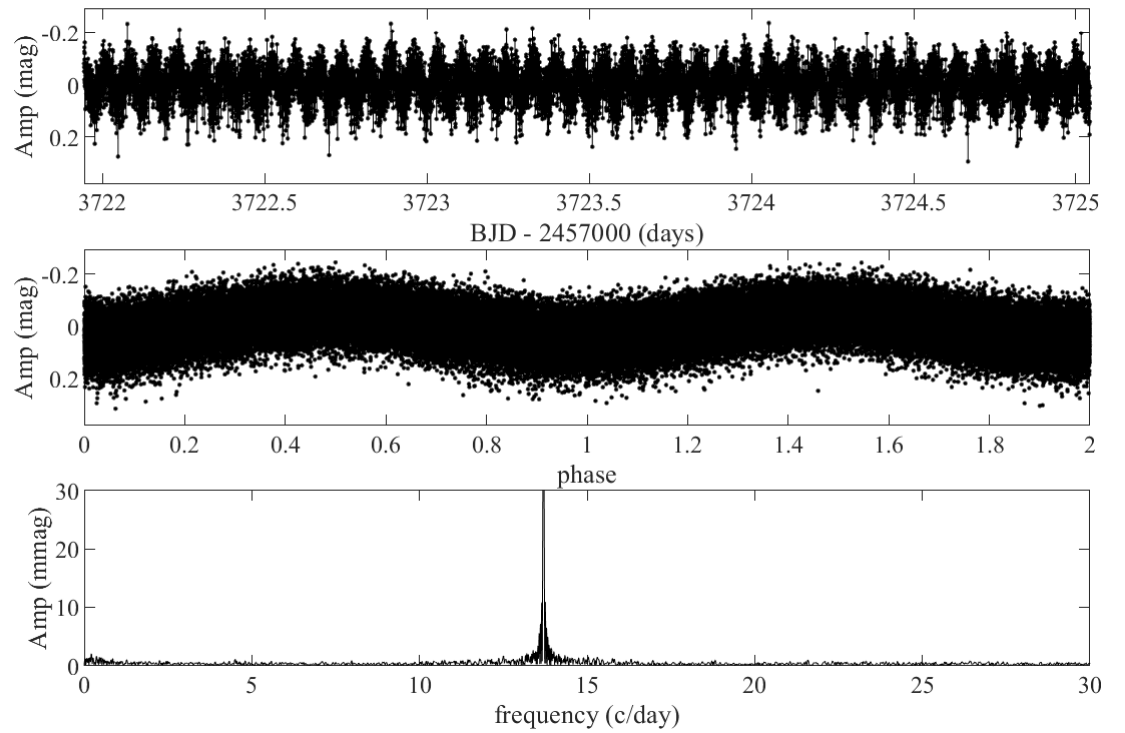}
\caption{The upper panel shows a portion of the light curve of TIC 36736263 for about 3 days; The middle panel shows the phase diagram folded by the only detected frequency $f = 13.69986$ c/day; The lower panel shows the Fourier amplitude spectra.}
\label{fig:TIC_36736263_light_curve}
\end{figure}

The light curve of TIC 36736263 exhibits a very symmetrical shape, resulting in only one frequency detected in the amplitude spectra, as shown in Figure \ref{fig:TIC_36736263_light_curve}. TIC 36736263 (another identifiers: XY Sex, PG1017-086), was discovered as a candidate of hot subdwarf in the Palomar-Green survey \citep{1986ApJS...61..305G}, and \cite{2002MNRAS.333..231M} derived its parameters of binary system (including the orbital period, mass function and projected rotational velocity) with photometric and spectroscopic observations and show that TIC 36736263 is an sdB star with a low-mass M-dwarf or brown dwarf companion.

\section{Period-Luminosity Relation} \label{sec:P-L relation}

The PL relation plays a dual role in astrophysics: it serves as a fundamental tool for calibrating cosmic distances \citep{1912HarCi.173....1L,1991PASP..103..933M}, while also providing critical insights into stellar physics and evolution \citep{2011ApJ...728L..43C,2025A&A...698A.304S}. To investigate this relation for our sample, we first computed the absolute magnitudes of the stars using

\begin{equation}\label{equ:ini_M10}
M_{V}=V - 5~\mathrm{log} ~ d + 5 - A_{V}
\end{equation}

where $M_{V}$ and $V$ denote the absolute and apparent magnitudes in the $V$ band, respectively, $d$ is the distance in pc, and $A_{V}$ is the extinction due to interstellar dust. Apparent magnitudes were obtained from TESS Input Catalog version 8.2 \citep[TIC v8.2,][]{2021arXiv210804778P}, accessed via VizieR\footnote{\url{https://vizier.cds.unistra.fr/viz-bin/VizieR-3}}. The stellar distances are from Gaia DR2 parallaxes, in which the normalized posterior distribution was used and length-scale model of \cite{2018AJ....156...58B} was adopted. For the extinctions and their uncertainties, we obtained their values from the 3D dust maps using the DUSTMAPS Python package \citep{2018JOSS....3..695G}, which provides access to the Bayestar 17 reddening map of \cite{2018MNRAS.478..651G}. 

\begin{figure}
\centering
\includegraphics[width=0.45\textwidth,trim=30 245 25 245,clip]{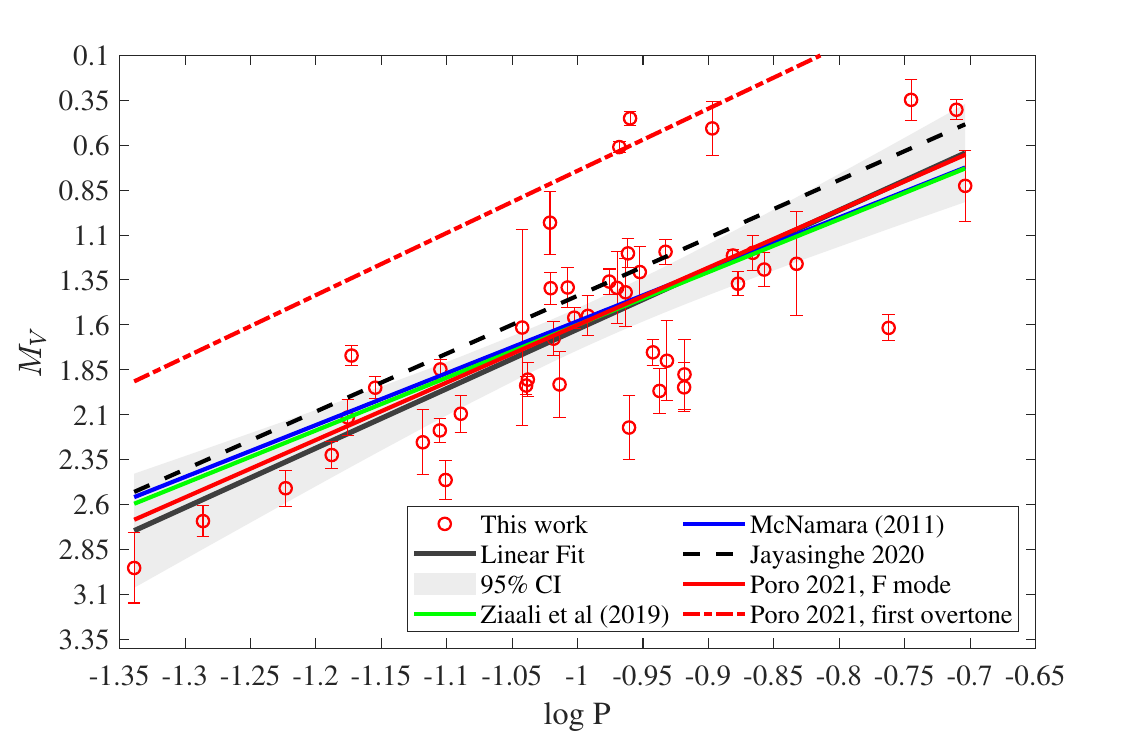}
\caption{PL relation of the 46 HADS stars in this work. The blue solid line is the PL relation of HADS stars derived by \cite{2011AJ....142..110M}, the green solid line is from \cite{2019MNRAS.486.4348Z}, the black dashed line is from \cite{2020MNRAS.493.4186J}, the red solid line (for the fundamental mode) and red dot-dashed line (for the first overtone mode) are from \cite{2021PASP..133h4201P}.}
\label{fig:Fig_p_L_relation}
\end{figure}

Figure \ref{fig:Fig_p_L_relation} shows the result of PL relation of the 46 HADS stars, as well as the relation by \cite{2011AJ....142..110M}, \cite{2019MNRAS.486.4348Z}, \cite{2020MNRAS.493.4186J}, and \cite{2021PASP..133h4201P}. With a sample of 46 HADS stars, we derived a new PL relation:

\begin{equation}\label{equ:P-L_relation}
M_{V}= (-3.31 \pm0.39)~\mathrm{log}~P+(-1.68 \pm 0.39)
\end{equation}

From Figure \ref{fig:Fig_p_L_relation}, it can be found that the new derived PL relation is in agreement with those by \cite{2011AJ....142..110M}, \cite{2019MNRAS.486.4348Z}, and \cite{2021PASP..133h4201P} for the fundamental mode. More than half of the stars (61\%) are concentrated within the 95\% confidence interval of the new relation, while several points in the middle of log $P$ deviate significantly. It is worth noting that three stars (TIC 78850814, TIC 30977864, and TIC 46937596) in our sample exhibit PL relations consistent with the first overtone mode, implying that their dominant frequencies may correspond to the first overtone. These targets should undergo detailed asteroseismic modelling in the future to verify their mode identities.

\section{Location in H-R Diagram} \label{sec:cite}

\begin{figure}
\centering
\includegraphics[width=0.45\textwidth,trim=10 10 10 20,clip]{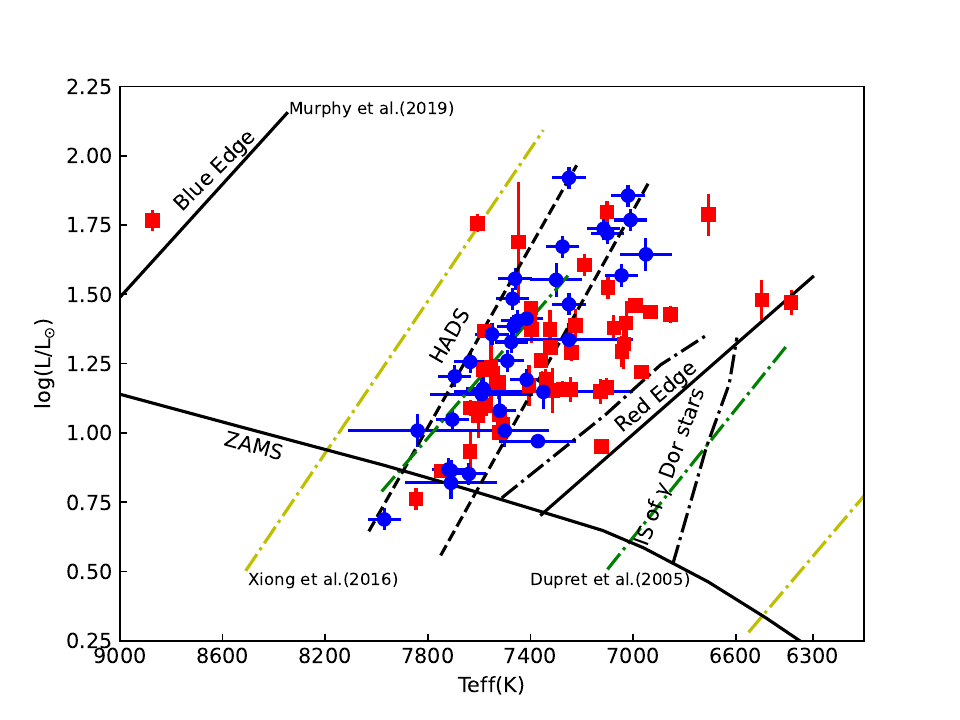}
\caption{Location of 46 HADS stars in H-R diagram in this work (red squares), along with 35 well-studied HADS (blue dots) collected from \cite{2000ASPC..210..373M,2005A&A...440.1097P,2011A&A...528A.147P,2007MNRAS.382..239C,2012MNRAS.419.3028B,2013MNRAS.433..394U,2016RMxAA..52..385P,2018ApJ...863..195Y,2019ApJ...879...59Y,2021MNRAS.504.4039B,2021A&A...655A..63Y}. The zero-age main-sequence (ZAMS) and the $\delta$ Sct instability strip (solid lines) are from \cite{2019MNRAS.485.2380M}. The yellow dot-dashed lines represent the boundaries of the instability strip predicted by \cite{2016MNRAS.457.3163X}. The dashed lines show the region occupied by HADS, as found by \cite{2000ASPC..210..373M}. The theoretical instability strip (IS) of $\gamma$ Dor stars (black dot-dashed lines), as well as the calculated positions of the blue and red edges for the fundamental mode (green dot-dashed lines), are from \cite{2005A&A...435..927D}.}
\label{fig:Fig_HRD}
\end{figure}

To investigate the evolutionary state of these 46 HADS stars, we calculated their luminosity ($L$) using: $M_{bol}=-2.5~log~(L/L_{\odot})+4.74$, where $L_{\odot}$ = 3.845$\times$ $10^{33}$ ergs s$^{-1}$, and +4.74 is the absolute bolometric magnitude of the Sun. The bolometric correction: $BC = M_{bol} - M_{V}$ was used by \cite{1981A&AS...46..193H}. The effective temperatures of these samples are obtained from TESS Input Catalog version 8.2 \citep[TIC v8.2,][]{2021arXiv210804778P}. Figure \ref{fig:Fig_HRD} displays the locations of the 46 HADS stars identified in this work, along with 35 well-studied HADS stars compiled from the literature \citep{2000ASPC..210..373M,2005A&A...440.1097P,2011A&A...528A.147P,2007MNRAS.382..239C,2012MNRAS.419.3028B,2013MNRAS.433..394U,2016RMxAA..52..385P,2018ApJ...863..195Y,2019ApJ...879...59Y,2021MNRAS.504.4039B,2021A&A...655A..63Y}. The figure also includes several theoretical and empirical instability strip boundaries for reference. The black solid lines represent the observationally determined blue and red edges from \cite{2019MNRAS.485.2380M}, derived from $\delta$~Sct stars observed by the \textit{Kepler} mission. The yellow dot-dashed lines indicate the theoretically predicted blue and red edges for low-order radial modes from \cite{2016MNRAS.457.3163X}, computed using a non-local and time-dependent treatment of convection. The green dot-dashed lines show the corresponding edges for the fundamental mode as calculated by \cite{2005A&A...435..927D}, who employed a time-dependent convection prescription.

From this figure, it can be found that while most sample HADS stars fall within the $\delta$ Sct instability strip of \cite{2019MNRAS.485.2380M} and \cite{2016MNRAS.457.3163X}, only one object lies beyond the blue edge of and another beyond the red edg \cite{2019MNRAS.485.2380M}. However, these stars are not fully concentrated in the narrow region of HADS stars reported by \cite{2000ASPC..210..373M}. In addition, based on the distance to the zero-age main-sequence, most of them are located in the main sequence, and a few may have already left the main sequence stage.

\section{Discussion} 

For the study of asteroseismology of pulsating stars, it is the first and crucial step to determine the number and identification of the modes. For $\delta$ Sct stars, their amplitude spectra are usually very rich and messy, which challenges the mode identifications. With high precision observations from TESS mission, \cite{2020Natur.581..147B} reported the detection of remarkably regular sequences of high-frequency pulsation modes in 60 $\delta$ Sct stars, which enables definitive mode identification. As a subclass of $\delta$ Sct star, HADS variable stars are generally considered to have very simple pulsation modes (the fundamental and first overtone mode), as there are always only one or two radial modes detected before space era. However, with the emergence of a large amount of high-precision time-series photometric data from space missions (such as MOST, CoRoT, Kepler, TESS), some low-amplitude frequencies are also detected, which form triplet or multiplet structures in amplitude spectra. These structures likely arise from distinct physical mechanisms, and their investigation offers valuable insights into stellar interiors and evolutionary process. 

In addition, the detection of low-amplitude frequencies in HADS stars raises important questions about pulsation properties. As photometric precision improves, we must reconsider whether truly mono-mode or double-mode HADS stars exist, or if all will eventually reveal non-radial modes. Our sample of 46 HADS stars currently contains apparent mono-mode and double-mode HADS stars (Table \ref{tab:list of single-mode HADS} and Table \ref{tab:list of pure double-mode HADS}), showing only fundamental or fundamental + first overtone modes in their amplitude spectra, with no significant residual peaks after frequencies removed. However, some marginal peaks below our significance threshold suggest that future higher-precision observations may reveal additional modes. This underscores the needs for future investigation into how pure radial pulsators differ from their non-radial counterparts in terms of internal structure, evolutionary stages, age, metallicity, etc.

In H-R diagram, compared to low amplitude $\delta$ Sct stars, HADS stars seems to lie in a narrow range at the center of the instability strip with a width of about 300 K in temperature \citep{2000ASPC..210..373M}, but observations from space photometry reveal that HADS stars do not seem to be located in any particular region of the instability strip\citep{2016MNRAS.459.1097B}. We plot a sample of 46 HADS stars identified in this work, as well as 35 well-studied HADS stars in H-R diagram, as shown in Figure \ref{fig:Fig_HRD}. From this figure, it can be found that while nearly all the HADS stars fall within the $\delta$ Sct instability strip, two stars lie beyond its boundaries. Notably, approximately half of the stars extend beyond the conversational red edge of the HADS region, suggesting cooler temperatures than typically associated with HADS variables. This distributions challenges the notion that a narrow 300 K instability strip uniquely distinguishes HADS stars from low amplitude $\delta$ Sct counterparts.

In recent years, the advent of space-based optical observation missions-most notably the ongoing TESS mission and the forthcoming PLATO\citep{2025ExA....59...26R}, CSST\citep{2025arXiv250704618C}, and ET2.0 \citep{2024ChJSS..44..400G} projects-has led to an unprecedented influx of high-precision photometric data. Although the primary scientific objectives of these missions are not primarily dedicated to variable star studies, the datasets they provide offer exceptional opportunities for the discovery and classification of variable stars. At the same time, the sheer volume of data poses significant challenges, making artificial intelligence (AI) an increasingly valuable tool for this task. Effective AI applications require large, accurately labeled training sets; insufficient or imprecise samples inevitably lead to misclassification or missed detections. For advancing HADS star searches, two critical needs emerge, including expansion of confirmed samples and improved classification of the sample. Our study addresses these needs by establishing a carefully classified HADS samples, creating an essential benchmark for future AI-assisted searches and accurate classification in large-scale photometric survey.

\section{Summary} 

In this work, we conduct a detailed pulsation analysis on a sample of HADS candidates with high-precision photometric data from TESS. All the significant frequencies are extracted in their amplitude and their dominant modes have been identified. Based on the number of the detected frequencies and the mode identifications, we find that 46 out of 50 stars belong to HADS stars, which are divided into 4 subgroups, including 7 pure mono-mode HADS stars (i.e. F mode), 21 HADS stars pulsating with the fundamental mode and non-radial modes (i.e. F + non-radial modes), 5 pure double-mode HADS stars (i.e. F+1O mode), 13 double-mode HADS with non-radial mode. The other 4 stars (TIC 1973623, TIC 8765832, TIC 32302937, and TIC 36736263) were misclassified as HADS stars before. Most of the dominant pulsation frequencies of the 7 pure mono-mode HADS stars are less than 10 c/day, while the frequency distribution of 21 HADS stars with non-radial pulsations is relatively wide, which may represent a difference between these two subgroups. In the double-mode HADS stars, 5 pure double-mode ones have a very narrow range of period ratio of F/1O (i.e. 0.77-0.78), while 8 out of 13 stars with non-radial modes lie in the range of 0.756 $\sim$ 0.787; In addition, 2 stars (TIC 33149129 and TIC 56882581) have the same period ratio of about 0.56, implying that they have the same non-radial modes; The other two stars (TIC 69546708 and TIC 110937533) also have the same period ratio of  about 0.80, which indicates they have the same modes. Among the 4 misclassified variable stars, we find that TIC 1973623 and TIC 36736263 are confirmed hot subdwarf stars and have been extensively studied, while TIC 32302937 is currently still a hot subluminous star candidates and need further spectral observations to verify its nature; TIC 8765832 is a cataclysmic variable and recent research shows that it lies close to the upper limit of the luminosity in which dwarf nova (DN) outbursts occur.

We investigated the PL relation of the 46 confirmed HADS variables and derived a revised formula: $M_{V}= (-3.31 \pm0.39)~\mathrm{log}~P+(-1.68 \pm 0.39)$, which agrees well with the result of previous studies. In H-R diagram, nearly all the HADS stars lie in the $\delta$ Sct instability strip; however, they are not confined to the narrow 300 K temperature range previously associated with HADS stars. Nearly half of our sample lies beyond the red edge of the HADS star region and tends to have lower temperatures. This distribution suggests that the distinction between HADS and low-amplitude $\delta$ Sct variables cannot be attributed solely to a specific location in the instability strip.

As upcoming space missions such as PLATO, CSST, and Earth 2.0 (ET) will generate unprecedented volumes of photometric data, the demand for efficient and accurate classification of variable stars increase dramatically. The high-quality HADS classifications achieved in this work not only provide an essential benchmark for future machine-learning-based searches and precise classifications, but also lay a solid foundation for advancing detailed asteroseismic modelling of these stars.

\begin{acknowledgments}
This research is supported by the National Natural Science Foundation of China (grant Nos. 12003020, 12473043) and Shaanxi Fundamental Science Research Project for Mathematics and Physics (Grant No. 23JSY015).

Machine-readable tables and light curves are provided in the online supplementary material.
\end{acknowledgments}

\begin{deluxetable}{rrrrc}
\tablewidth{0pt}
\tablecaption{The Detected Frequencies in Short Cadence (SC) Data of TIC 710783 
\label{tab:TIC710783}}
\tablehead{
\colhead{$f_{i}$} & \colhead{Frequency } & \colhead{Amplitude } & \colhead{S/N} & \colhead{Comment} \\ 
\colhead{} & \colhead{ (c/day) } & \colhead{ (mmag) }
}
\startdata
 1 &  12.74450 $\pm$ 0.00002 &  168.97 $\pm$ ~ 0.36 &809.8   & F0    \\
 2 &  25.48902 $\pm$ 0.00002 &   64.41 $\pm$ ~ 0.18 &601.5   & 2F0   \\
 3 &  38.23343 $\pm$ 0.00004 &   25.38 $\pm$ ~ 0.14 &316.3   & 3F0   \\
 4 &  50.97777 $\pm$ 0.00008 &   12.52 $\pm$ ~ 0.12 &176.9   & 4F0   \\
 5 &  63.72247 $\pm$ 0.00016 &    6.18 $\pm$ ~ 0.12 & 85.3   & 5F0   \\
 6 &  76.46695 $\pm$ 0.00029 &    3.69 $\pm$ ~ 0.14 & 46.7   & 6F0   \\
 7 &  46.93186 $\pm$ 0.00078 &    1.29 $\pm$ ~ 0.13 & 17.3   & nonradial   \\
 8 &  59.67568 $\pm$ 0.00086 &    1.16 $\pm$ ~ 0.13 & 15.6   & F0+f7   \\
 9 &  72.41862 $\pm$ 0.00109 &    0.84 $\pm$ ~ 0.12 & 12.3   & 2F0+f7 \\
10 &  11.83773 $\pm$ 0.00244 &    0.81 $\pm$ ~ 0.25 &  5.5   & nonradial   \\
\enddata
\tablecomments{The strongest peak is the fundamental mode, others are harmonics, combinations and nonradial mode.}
\tablecomments{Table 6 is published in its entirety in the electronic edition of the {\it Astrophysical Journal}. Just the short cadence data of TIC 710783 portion is shown here for guidance regarding its form and content. The full machine readable version has the data for all 50 sources.}
\end{deluxetable}

\bibliography{sample7}{}
\bibliographystyle{aasjournalv7}

\end{document}